\documentclass[english,pra,showpacs,showkeys,tightenlines,secnumarabic,11pt]{revtex4}
\usepackage[T1]{fontenc}
\usepackage[latin1]{inputenc}
\usepackage{amsmath}
\usepackage{graphicx}
\usepackage{amssymb}
\usepackage{epsfig}
\usepackage{graphics}
\usepackage[mathscr]{euscript}
\usepackage{psfrag}
\usepackage{pstricks}
\usepackage{pst-node}
\usepackage{babel}

\begin{document}
\title{About Double Parton Scattering at Short Relative Transverse Distances}
\author{D. Treleani}\email{daniele.treleani@ts.infn.it; Retired} \affiliation{ Dipartimento di
Fisica dell'Universit\`a di Trieste and INFN, Sezione di
Trieste,\\ Strada Costiera 11, Miramare-Grignano, I-34151 Trieste,
Italy.}

\author{G. Calucci}
\email{giorgio.calucci@alice.it; Retired} \affiliation{ Dipartimento di
Fisica dell'Universit\`a di Trieste and INFN, Sezione di
Trieste,\\ Strada Costiera 11, Miramare-Grignano, I-34151 Trieste,
Italy.}

\begin{abstract}
By working out the kinematics of double parton scattering at short relative transverse distances, we obtain an explicit link between the transverse centres of mass, of the two hard partonic interactions, and the contributions to the process, due to pairs of interacting partons generated by perturbative splitting. One my thus foresee the interesting possibility of discriminating experimentally between contributions to the double parton scattering cross section, due to interacting parton pairs originated by independent evolution, and contributions, due to interacting parton pairs generated by splitting.
\end{abstract}



\maketitle

\section{Introduction}

In hadronic collisions double parton scattering (DPS) plays an increasingly important role at high energies\cite{Bartalini,Aaij:2015wpa,Diehl:2011yj,Belyaev:2017sws,Aaboud:2016dea,Sirunyan:2017hlu,Lansberg:2017chq,Maciula:2018mig,Kasemets:2017vyh,Traini:2016jru,dEnterria:2016yhy}. In DPS the hard component of the interaction is factorised in two components, each involving two different pairs of initial state partons. The process is thus described by two scales: the large scale, represented by the squared momenta transferred in the hard interactions ${\cal Q}^2$, and the small scale, given by the soft relative momentum between the two hard interactions. The cross section is derived in the limiting case where the small scale can be neglected,  whenever possible, as compared with ${\cal Q}^2$, or it is integrated over\cite{Bartalini}\cite{Diehl:2011yj}\cite{Paver:1982yp}. The integration on the small scale is conveniently transformed into an integration over the relative transverse distance between the positions of the two interactions, which are localised in two different points, within the overlap volume of the two colliding hadrons. The non-perturbative component of the process is hence factorised as a product of two double parton distributions (DPDs), which depend on ${\cal Q}^2$, on the fractional momenta of the initial state partons $x_i$ and on their relative transverse distance $b$. At the leading order in ${\cal Q}^2$, the latter quantity is integrated over and therefore unobservable in the final state. In this way, in addition to the dependence on initial fractional momenta, on ${\cal Q}^2$, and on the final momenta at large $p_t$, the DPS cross section depends also on a non-perturbative quantity, resulting from the integration of the two DPDs on $b$, namely the effective cross section $\sigma_{eff}$. 

On the other hand in a DPS one can measure also the transverse momenta of the centres of mass of the two hard interactions, which, although sizeably smaller as compared with ${\cal Q}$, are large enough to be discussed in perturbation theory. Including explicitly the dependence of the DPS process on the transverse momenta of the c.m. of the two hard partonic interactions, is however a non trivial multi-scale problem in pQCD\cite{Diehl:2011yj}\cite{Blok:2011bu,Blok:2013bpa,Diehl:2017kgu,Buffing:2017mqm,Diehl:2017wew}, whose exhaustive study requires a considerable effort. 

 To the purpose of obtaining indications on possible hidden features of the process, which might be disclosed taking advantage of the additional information provided by a less inclusive cross section, we think helpful to explore a simplified case. In the present note, we therefore work out the explicit dependence of the process, as a function of the transverse momenta of the c.m. of the two hard partonic interactions, in the simplest instance of DPS at the lowest order in the coupling constant.

The DPS amplitude includes a loop integral\cite{Diehl:2011yj}\cite{Paver:1982yp}, whose range is limited by the hadron form factor and which plays a relevant role, when working out the dependence of the initial partonic state on the c.m. of the two hard interactions. As it will be shown hereafter, the DPS production rate is in fact linked, through the loop, to contributions to the DPS process at short relative transverse distances, the dominant contribution being perturbative splitting (e.g. the $3\to4$ parton processes\cite{Blok:2011bu}\cite{Blok:2013bpa}).

The paper is organised as follows: the next section, mainly devoted to kinematics, is divided in two sub-sections. In the first sub-section we remind the steps to obtain the DPDs in the usual kinematics of DPS. In the following sub-section we work out the kinematics to obtain the DPDs, without integrating on the c.m. transverse momenta of the two hard collisions, and we point out some related interesting features of the process. The last section is devoted to the concluding discussion.

\section{Kinematics of DPS}

\subsection{Double Parton Distributions}

\par
\begin{figure}[h]
\centering
\includegraphics[width=13cm]{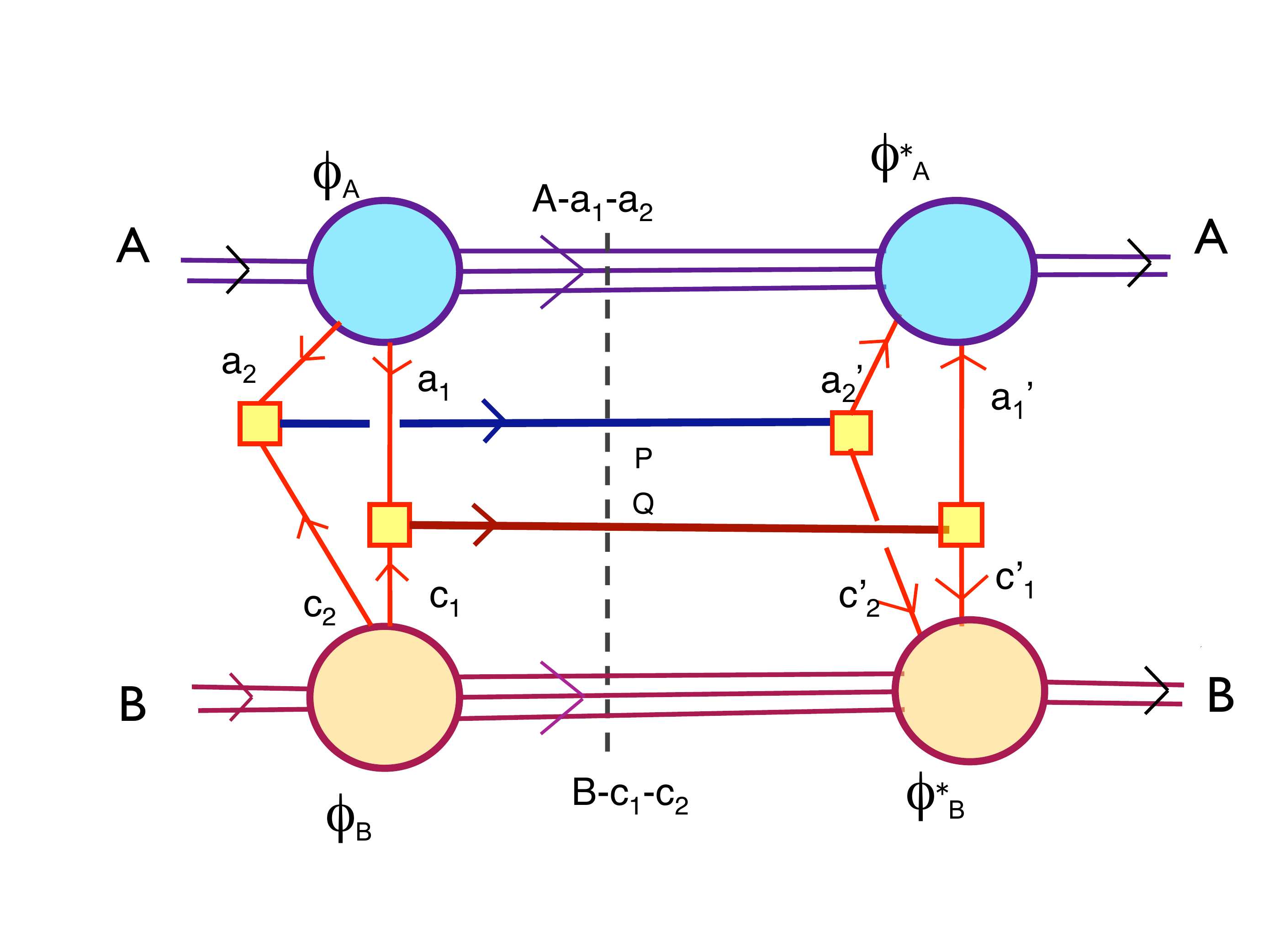}
\caption{DPS contribution to the forward amplitude}
\label{DPS1}
\end{figure}

\begin{figure}[h]
\centering
\includegraphics[width=13cm]{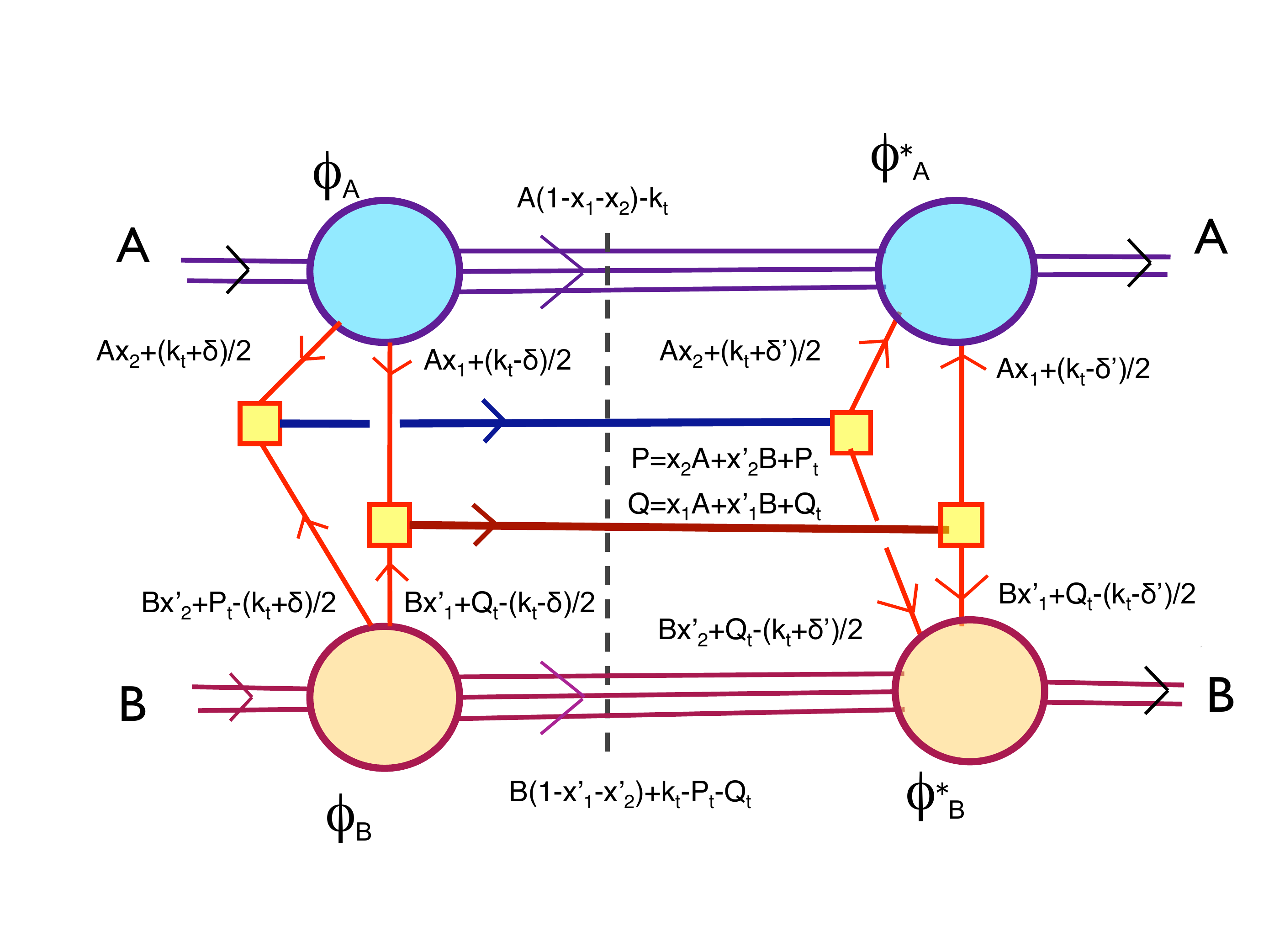}
\caption{Flow of momenta in the DPS diagram}
\label{DPS2}
\end{figure}

 The DPS contribution to the forward amplitude is shown in Fig.\ref{DPS1}, while all momenta are explicitly indicated in Fig.\ref{DPS2}.

\par To evaluate the leading contribution to the DPS cross section only the longitudinal momentum components which grow as $\sqrt{s}$ are taken into account. In this way the integrations on the light cone components $\delta_-$, $\delta_-'$ and $\delta_+$, $\delta_+'$ involve only the upper and the lower parts respectively of the diagrams in Figures \ref{DPS1} and \ref{DPS2}. One may thus define the quantities

\begin{eqnarray}
&&\Psi_A(x_i,k_t,\bold\delta_t)=\frac{1}{\sqrt 2}\int \frac{{\Phi}_A}{a_1^2a_2^2}\frac{d\delta_-}{2\pi},\qquad\Psi_B(x_i',k_t,\bold\delta_t)=\frac{1}{\sqrt 2}\int \frac{{\Phi}_B}{c_1^2c_2^2}\frac{d\delta_+}{2\pi}\cr
&&\cr
&&\Psi_A^*(x_i,k_t,\bold\delta_t')=\frac{1}{\sqrt 2}\int \frac{{\Phi}_A^*}{a_1'^2a_2'^2}\frac{d\delta_-'}{2\pi},\qquad\Psi_B^*(x_i',k_t,\bold\delta_t')=\frac{1}{\sqrt 2}\int \frac{{\Phi}_B^*}{c_1'^2c_2'^2}\frac{d\delta_+'}{2\pi}
\end{eqnarray}
	
The fractional momentum components $x_i$, $x'_i$ are fixed by the final state partons with large transverse momenta while, to evaluate the cross section, the transverse momentum components $\bold a_{i,t}$, $\bold a_{i,t}'$, $\bold c_{i,t}$, $\bold c_{i,t}'$ have to be integrated. To this purpose one may introduce the Fourier transforms 

\begin{eqnarray}
\tilde\Psi_A(x_i,b_1,b_2)=\frac{1}{(2\pi)^2}\int e^{i\bold{b}_1\cdot\bold{a}_{1,t}}e^{i\bold{b}_2\cdot\bold{a}_{2,t}}\Psi_A(x_i,a_{1,t},a_{2,t})d^2a_{1,t}d^2a_{2,t}\qquad{\rm etc.}
\end{eqnarray}

Expressing $\bold a_{i,t},\,\bold a_{i,t}',\,\bold c_{i,t},\,\bold c_{i,t}'$ in terms of the independent degrees of freedom $\bold k_t,\,\boldsymbol\delta_t,\bold P_t,\,\boldsymbol\delta_t',\,\bold Q_t$ as shown in Fig.\ref{DPS2} (e.g. $\bold{a}_{1,t}=(\bold k_t-\boldsymbol\delta_t)/2,\, \bold c_{1,t}=\bold Q_t- (\bold k_t-\boldsymbol\delta_t)/2$, and so on) the integrations on the transverse variables are 

\begin{eqnarray}
&&\frac{1}{(2\pi)^{10}}\int\tilde\Psi_A(b_1,b_2)\tilde\Psi_B(b_3,b_4)\tilde\Psi_A^*(b_1',b_2')\tilde\Psi_B^*(b_3',b_4')\,\,d^2b_1d^2b_2d^2b_3d^2b_4d^2b_1'd^2b_2'd^2b_3'd^2b_4'd^2\delta_t d^2\delta_t'\cr
&&\quad\times{\rm exp}\Bigl[i\Bigl\{\frac{1}{2}(\bold k_t+\boldsymbol\delta_t)\cdot\bold b_1+\frac{1}{2}(\bold k_t-\boldsymbol\delta_t)\cdot\bold b_2+\big[\frac{1}{2}(-\bold k_t-\boldsymbol\delta_t)+\bold P_t\big]\cdot\bold b_3+\big[\frac{1}{2}(-\bold k_t+\boldsymbol\delta_t)+\bold Q_t\big]\cdot\bold b_4\cr
&&\quad-\frac{1}{2}(\bold k_t+\boldsymbol\delta_t')\cdot\bold b_1'-\frac{1}{2}(\bold k_t-\boldsymbol\delta_t')\cdot\bold b_2'-\big[\frac{1}{2}(-\bold k_t-\boldsymbol\delta_t')+\bold P_t\big]\cdot\bold b_3'-\big[\frac{1}{2}(-\bold k_t+\boldsymbol\delta'_t)+\bold Q_t\big]\cdot\bold b_4'\Bigr\}\Big]
\label{loop_int}
\end{eqnarray}

\noindent
where the dependence on $x_i,\,x_i'$ is understood. By integrating on $\boldsymbol\delta_t$ and $\boldsymbol\delta_t'$ one obtains

\begin{eqnarray}
d^2\delta_t\to(2\pi)^2\delta(\bold b_1-\bold b_2-\bold b_3+\bold b_4),\qquad d^2\delta_t'\to(2\pi)^2\delta(\bold b_1'-\bold b_2'-\bold b_3'+\bold b_4')
\end{eqnarray}

\noindent
which implies

\begin{eqnarray}
\bold b_1-\bold b_2=\bold b_3-\bold b_4\equiv \bold b,\qquad \bold b_1'-\bold b_2'=\bold b_3'-\bold b_4'\equiv\bold b'
\end{eqnarray}

\noindent
where $\bold b$ and $\bold b'$ are the relative distance between the two interactions, in the amplitude of the process and in its complex conjugate. Only fragments with large $p_t$ are observed in the final state. One can therefore integrate also on $\bold k_t$:

\begin{eqnarray}
d^2k_t\to(2\pi)^2\delta(\bold b_1+\bold b_2-\bold b_3-\bold b_4-\bold b_1'-\bold b_2'+\bold b_3'+\bold b_4')
\label{kt_int}
\end{eqnarray}

By making the positions

\begin{eqnarray}
\frac{1}{2}(\bold b_1+\bold b_2)\equiv\bold B_1,\qquad\frac{1}{2}(\bold b_3+\bold b_4)\equiv\bold B_3,\qquad\frac{1}{2}(\bold b_1'+\bold b_2')\equiv\bold B_1',\qquad\frac{1}{2}(\bold b_3'+\bold b_4')\equiv\bold B_3'
\end{eqnarray}

\noindent
and using the constraint ($\ref{kt_int}$), one obtains the following relations between the c.m. coordinates $\bold B_i$, $\bold B_i'$:

\begin{eqnarray}
\bold B_1-\bold B_3=\bold B_1'-\bold B_3'\equiv\boldsymbol\Delta\quad\Rightarrow\quad \bold B_3=\bold B_1-\boldsymbol\Delta,\,\,\bold B_3'=\bold B_1'-\boldsymbol\Delta
\end{eqnarray}

Expression ($\ref{loop_int}$) thus simplifies to

\begin{eqnarray}
&&\frac{1}{(2\pi)^4}\int\tilde\Psi_A(b,B_1)\tilde\Psi_B(b,B_1-\Delta)\tilde\Psi_A^*(b',B_1')\tilde\Psi_B^*(b',B_1'-\Delta)d^2bd^2b'd^2B_1d^2B_1'd^2\Delta\cr
&&\qquad\times{\rm exp}\Bigl[i\Bigl\{\bold P_t\cdot\big(\bold B_3+\frac{\bold b}{2}\big)+\bold Q_t\cdot\big(\bold B_3-\frac{\bold b}{2}\big)-\bold P_t\cdot\big(\bold B_3'+\frac{\bold b'}{2}\big)-\bold Q_t\cdot\big(\bold B_3'-\frac{\bold b'}{2}\big)\Bigr\}\Big]
\end{eqnarray}

The argument of the exponential is

\begin{eqnarray}
&\bigl\{(\bold P_t+\bold Q_t)\cdot(\bold B_3-\bold B_3')+\frac{1}{2}(\bold P_t-\bold Q_t)\cdot(\bold b-\bold b')\bigr\}\cr
=&\bigl\{(\bold P_t+\bold Q_t)\cdot(\bold B_1-\bold B_1')+\frac{1}{2}(\bold P_t-\bold Q_t)\cdot(\bold b-\bold b')\bigr\}
\end{eqnarray}

\noindent
and the transverse integrations are 

\begin{eqnarray}
&&\frac{1}{(2\pi)^4}\int d^2\Delta\int\tilde\Psi_A(b,B_1)\tilde\Psi_B(b,B_1-\Delta)\times e^{i(\bold P_t+\bold Q_t)\cdot\bold B_1}e^{i(\bold P_t-\bold Q_t)\cdot\bold b/2}d^2B_1d^2b\cr
&&\qquad\qquad\,\times\tilde\Psi_A^*(b',B_1')\tilde\Psi_B^*(b',B_1'-\Delta)\times e^{-i(\bold P_t+\bold Q_t)\cdot\bold B_1'}e^{-i(\bold P_t-\bold Q_t)\cdot\bold b'/2}d^2B_1'd^2b'\cr
&&\cr
&&=\int d^2\Delta\Big|\frac{1}{(2\pi)^2}\int\tilde\Psi_A(b,B_1)\tilde\Psi_B(b,B_1-\Delta)\times e^{i(\bold P_t+\bold Q_t)\cdot\bold B_1}e^{i(\bold P_t-\bold Q_t)\cdot\bold b/2}d^2B_1d^2b\Big|^2
\label{Kintegrated}
\end{eqnarray}

After integrating on $\bold P_t$ and $\bold Q_t$, one thus obtains 

\begin{eqnarray}
&&\int |\tilde\Psi_A(x_i,b,B_1)|^2|\tilde\Psi_B(x_i',b,B_1-\Delta)|^2d^2B_1d^2bd^2\Delta\cr
&&\qquad\qquad\qquad=\int d^2b\int|\tilde\Psi_A(x_i,b,B_1)|^2d^2B_1\int|\tilde\Psi_B(x_i',b,B_1')|^2d^2B_1'
\label{usual}
\end{eqnarray}

\noindent
where the dependence on the fractional momenta $x_i$ is explicitly indicated. Multiplying $\int|\tilde\Psi_A(x_i,b,B_1)|^2d^2B_1$ and $\int|\tilde\Psi_B(x_i',b,B_1')|^2d^2B_1'$ by the proper $x_i$-dependent factors, namely the flux factors of the elementary partonic cross sections and the factors deriving from the integration on the invariant mass of the residual hadron fragments, one obtains the DPDs\cite{Calucci:2010wg}. 

By integrating on $\bold P_t$ and $\bold Q_t$ any explicit connection of the transverse distance $b$ with the final state produced is therefore lost and, in this way, $b$ has become a hidden degree of freedom in the process. 

The final, non straightforward, step to obtain the non-perturbative input to the DPS cross section is to evolve the DPDs up to the large scale ${\cal Q}^2$, of the two hard interactions\cite{Diehl:2011yj}\cite{Blok:2011bu,Blok:2013bpa,Diehl:2017kgu,Buffing:2017mqm,Diehl:2017wew}\cite{Ryskin:2011kk,Manohar:2012pe,Elias:2017flu}.

\subsection{Double Parton Distributions as a function of $\bold P_t$ and $\bold Q_t$}

A connection between the final state produced by DPS and the relative transverse distance between the two hard interactions can be obtained, by keeping alive the explicit dependence of the initial state configuration on $\bold P_t$ and on $\bold Q_t$. To this aim it is convenient to use the mixed representation of the states, $\bar\Psi$, defined by the Fourier transforms

\begin{eqnarray}
\bar\Psi_A(b,K_1)=\frac{1}{2\pi}\int \tilde\Psi_A(b,B_1)e^{i\bold K_1\cdot \bold B_1}d^2B_1,\qquad{\rm etc.}
\end{eqnarray}

Expression (\ref{Kintegrated}) is thus written as

\begin{eqnarray}
&&\frac{1}{(2\pi)^8}\int\bar\Psi_A(b,K_1)\bar\Psi_B(b,K_2)\bar\Psi_A^*(b',K_1')\bar\Psi_B^*(b',K_2')\,\,d^2B_1d^2B_1'd^2bd^2b'd^2\Delta d^2K_1d^2K_2d^2K_1'd^2K_2'\cr
&&\qquad\qquad\times {\rm exp}\Big\{i\big[(\bold P_t+\bold Q_t)\cdot\bold B_1+\bold K_1\cdot\bold B_1+\bold K_2\cdot(\bold B_1-\boldsymbol\Delta)+(\bold P_t-\bold Q_t)\cdot\bold b/2\cr
&&\qquad\qquad\qquad\quad-(\bold P_t+\bold Q_t)\cdot\bold B_1'-\bold K_1'\cdot\bold B_1'-\bold K_2'\cdot(\bold B_1'-\boldsymbol\Delta)-(\bold P_t-\bold Q_t)\cdot\bold b'/2\big]\Big\}
\label{FKintegrated}
\end{eqnarray}

\noindent
which allows integrating explicitly on $\boldsymbol\Delta$, $\bold B_1$ and $\bold B_1'$. One obtains

\begin{eqnarray}
&&d^2\Delta\to(2\pi)^2\delta(-\bold K_2+\bold K_2')\cr 
&&d^2B_1\to(2\pi)^2\delta(\bold P_t+\bold Q_t+\bold K_1+\bold K_2)\cr 
&&d^2B_1'\to(2\pi)^2\delta(\bold P_t+\bold Q_t+\bold K_1'+\bold K_2')
\end{eqnarray}

\noindent
namely

\begin{eqnarray}
\bold K_2=-\bold K_1-\bold P_t-\bold Q_t,\qquad \bold K_2'=-\bold K_1'-\bold P_t-\bold Q_t,\qquad \bold K_2=\bold K_2',\qquad \bold K_1=\bold K_1'
\end{eqnarray}

Expression(\ref{FKintegrated}) thus becomes

\begin{eqnarray}
&&\,\,\,\,\,\,\frac{1}{(2\pi)^2}\int\bar\Psi_A(b,K_1)\bar\Psi_A^*(b',K_1)\bar\Psi_B(b,-K_1-P_t-Q_t)\bar\Psi_B^*(b',-K_1-P_t-Q_t)\cr
&&\qquad\qquad\qquad\qquad\qquad\qquad\qquad\qquad\times e^{i(\bold P_t-\bold Q_t)\cdot(\bold b-\bold b')/2}d^2bd^2b'd^2K_1\cr
&&=\frac{1}{(2\pi)^2}\int d^2K_1\int\bar\Psi_A(b,K_1)\bar\Psi_B(b,-K_1-P_t-Q_t)\,e^{i(\bold P_t-\bold Q_t)\cdot\bold b/2}\,d^2b\cr
&&\,\,\qquad\qquad\qquad\times\,\int\bar\Psi_A^*(b',K_1)\bar\Psi_B^*(b',-K_1-P_t-Q_t)\,e^{-i(\bold P_t-\bold Q_t)\cdot\bold b'/2}\,d^2b'\cr
&&=\int d^2K_1\Big|\frac{1}{2\pi}\int\bar\Psi_A(b,K_1)\bar\Psi_B(b,-K_1-P_t-Q_t)\,e^{i(\bold P_t-\bold Q_t)\cdot\bold b/2}\,d^2b\Big|^2
\label{NFKsub}
\end{eqnarray}

As pointed out by several authors\cite{Diehl:2011yj}\cite{Blok:2011bu,Blok:2013bpa}\cite{Manohar:2012pe}, at short relative transverse distances, the dominant source of the two partons, undergoing the hard interactions, is perturbative splitting, which induces a singular behaviour in the DPDs. On the other hand, the singular behaviour of the DPDs at short transverse distances is a manifestation of the transition of the interaction from a double to a single parton scattering (SPS) process\cite{Diehl:2011yj}. To be properly defined and to avoid double counting with SPS, the singular term at short relative transverse distances has thus to be subtracted in the definition of the DPS cross section\cite{Blok:2011bu}\cite{Diehl:2017kgu}.

One should thus consider the case where $\bar\Psi(b,K)$ is regular as a function of $K$, while it goes as $1/b$ for $b$ small\cite{Diehl:2011yj}\cite{Diehl:2017kgu}. To simplify the discussion, we consider the following factorised expression:

\begin{eqnarray}
\bar\Psi(x_i,b,K_1)=\Big[\frac{\psi(x_i,b)}{b}+\eta(x_i,b)\Big]\varphi(x_i,K_1)
\end{eqnarray}

Here the dependence on the fractional momenta $x_i$, $x_i'$ is explicitly indicated and we assume $\psi(x_i,b)$, $\eta(x_i,b)$  and $\varphi(x_i,K_1)$ to be regular functions of $b$ and of $K_1$ respectively\footnote{On general grounds, $\eta(x_i,b)$ can have a logarithmic singularity as a function of $b$, which however would modify only marginally the actual discussion.}. Expression (\ref{NFKsub}) is therefore given by

\begin{eqnarray}
&&\frac{1}{(2\pi)^2}\int d^2K_1\Big|\int d^2b\Big[\frac{\psi_A(x_i,b)}{b}+\eta_A(x_i,b)\Big]\Big[\frac{\psi_B(x_i',b)}{b}+\eta_B(x_i',b)\Big]\,e^{i(\bold P_t-\bold Q_t)\cdot\bold b/2}\cr
&&\qquad\qquad\qquad\qquad\qquad\qquad\qquad\qquad\quad\qquad\times\varphi_A(x_i,K_1)\varphi_B(x_i',-K_1-P_t-Q_t)\Big|^2\cr
=&&\Big|\frac{1}{2\pi}\int d^2b\Big[\frac{\psi_A(x_i,b)}{b}+\eta_A(x_i,b)\Big]\Big[\frac{\psi_B(x_i',b)}{b}+\eta_B(x_i',b)\Big]\,e^{i(\bold P_t-\bold Q_t)\cdot\bold b/2}\Big|^2\cr
&&\qquad\qquad\qquad\qquad\qquad\qquad\qquad\quad\times\int d^2K_1\Big|\varphi_A(x_i,K_1)\varphi_B(x_i',-K_1-P_t-Q_t)\Big|^2
\label{model}
\end{eqnarray}

The integrations on $\bf b$ and on $\bold K_1$ are independent one to another. When $\bold P_t-\bold Q_t$ is finite, the integral on $\bf b$ is logarithmically divergent for $b\to 0$ and needs to be regularised. The integration on $\bold K_1$ does not have problems and will be done as first. To proceed we consider the simplest case of a gaussian distribution of partons within a hadron of radius R:

\begin{eqnarray}
\varphi_A(x_i,K_1)=h_A(x_i)\frac{R}{\sqrt {2\pi}}\,e^{-K_1^2R^2/4}\quad{\rm etc.}
\end{eqnarray}

One obtains

\begin{eqnarray}
\int d^2K_1\Big|\varphi_A(x_i,K_1)\varphi_B(x_i',-K_1-P_t-Q_t)\Big|^2= \big|h_A(x_i)h_B(x_i')\big|^2\frac{R^2}{4\pi}e^{-|\bold P_t+\bold Q_t|^2R^2/4}
\end{eqnarray}

The angular integration on $\bf b$ involves only the exponential and the result is the Bessel function of the first kind $J_0$. The radial integration is 

\begin{eqnarray}
\int_{b_{min}}^{\infty}\Big[\frac{\psi_A(x_i,b)}{b}+\eta_A(x_i,b)\Big]\Big[\frac{\psi_B(x_i',b)}{b}+\eta_B(x_i',b)\Big]\,J_0(|\bold P_t-\bold Q_t|b/2)\,bdb
\label{radial}
\end{eqnarray}

\noindent
where $b_{min}\equiv 1/{\cal S}$ is a lower cutoff, introduced to regularise the integral at small $b$ and we consider only events where $|\bold P_t-\bold Q_t|<{\cal S}$. As a simplest model one may take

\begin{eqnarray}
\psi_{A,B}(x_i,b)=f(x_i)\frac{e^{-b^2/(4R^2)}}{\sqrt{2\pi}},\qquad\eta_{A,B}(x_i,b)=g(x_i)\frac{e^{-b^2/(4R^2)}}{\sqrt{2\pi} R}
\end{eqnarray}

\noindent
which gives 

\begin{eqnarray}
\int_{b_{min}}^{\infty}\Big[\frac{f(x_i)}{b}+\frac{g(x_i)}{ R}\Big]\Big[\frac{f(x_i')}{b}+\frac{g(x_i')}{ R}\Big]\,e^{-b^2/(2R^2)}\,J_0(|\bold P_t-\bold Q_t|b/2)\,bdb
\end{eqnarray}
 
The integral on $b$ has therefore three different contributions:

\begin{eqnarray}
C_1&=&\int_0^{\infty}\frac {e^{-b^2/(2R^2)}}{b}\big[1-J_0(|\bold P_t-\bold Q_t|b/2)\big]db=\frac{1}{2}\Big[\gamma-{\rm Ei}\Big(-\frac{|\bold P_t-\bold Q_t|^2R^2}{2}\Big)+{\rm ln}\Big(\frac{|\bold P_t-\bold Q_t|^2R^2}{2}\Big)\Big]\cr
C_2&=&\int_0^{\infty}\frac{1}{ R}e^{-b^2/(2R^2)}J_0(|\bold P_t-\bold Q_t|b/2)db=\sqrt\frac{\pi}{2}\,\, e^{-|\bold P_t-\bold Q_t|^2R^2/4}\,I_0
\Big(\frac{|\bold P_t-\bold Q_t|^2R^2}{4}\Big)\cr
C_3&=&\int_0^{\infty}\frac{1}{ R^2}e^{-b^2/(2R^2)}J_0(|\bold P_t-\bold Q_t|b/2)bdb=e^{-|\bold P_t-\bold Q_t|^2R^2/2}
\label{TDC}
\end{eqnarray}

\noindent
where $\gamma$ is the Euler-Mascheroni constant, ${\rm Ei}$ the exponential integral and $I_0$ the modified Bessel function of the first kind. The singularity of $C_1$, for $b_{min}\to0$, has been removed by subtracting from the Bessel function $J_0$ its value at the origin, while the lower limit of the integral has been extended to 0. 

The three contributions have a very different dependence on $|\bold P_t-\bold Q_t|$: $C_1$ vanishes for $|\bold P_t-\bold Q_t|\to 0$, it becomes sizeable when $|\bold P_t-\bold Q_t|\simeq 2/R$ and it grows logarithmically when $|\bold P_t-\bold Q_t|>2/R$. $C_2$ and $C_3$ assume, on the contrary, their maximal value for $|\bold P_t-\bold Q_t|=0$ and decrease rapidly to zero for $|\bold P_t-\bold Q_t|$ greater than $2/R$. When $|\bold P_t-\bold Q_t|<2/R$, expression (\ref{radial}) is dominated by the sum $C_2+C_3$, while it coincides practically with $C_1$ when $|\bold P_t-\bold Q_t|>2/R$. 

One can therefore identify two different regions, which do not interfere with one another. The first region, $|\bold P_t-\bold Q_t|<2/R$, is the region where all initial state partonic configurations are non-perturbative, including interference terms with configurations where pairs of interacting partons are generated non-perturbatively. In this region $\bold P_t$ and $\bold Q_t$ are either very small or almost parallel and close in length, their difference being at most $2/R\approx 500\,{\rm MeV}$. 

In the second region, $|\bold P_t-\bold Q_t|>2/R$, the interacting parton pairs are generated by perturbative splitting. For large $|\bold P_t-\bold Q_t|$ the population of parton pairs is given by:

\begin{eqnarray}
&&\Big|f(x_i)f(x_i')\frac{1}{2}\big[\gamma-{\rm ln}2+{\rm ln}\big(|\bold P_t-\bold Q_t|^2R^2\big)\big]\Big|^2\cr
&&\qquad\qquad\qquad\qquad\to\frac{1}{2}|f(x_i)|^2{\rm ln}\big(|\bold P_t-\bold Q_t|^2R^2\big)\times\frac{1}{2}|f(x_i')|^2{\rm ln}\big(|\bold P_t-\bold Q_t|^2R^2\big)
\label{popul}
\end{eqnarray}

The growth of the double parton distributions at small relative transverse distances is thus given, both for the projectile and for the target separately, by the logarithmic evolution from the initial hadronic scale, $1/R^2$, to the "large scale", $|\bold P_t-\bold Q_t|^2$. Notice that this "small range" evolution is different in each event and, for $|\bold P_t|$ and $|\bold Q_t|$ fixed, it is maximal when $\bold P_t$ and $\bold Q_t$ are back to back. 

The DPS cross section depends therefore (logarithmically) on $|\bold P_t-\bold Q_t|$, whose upper bound is the arbitrary scale ${\cal S}\equiv1/b_{min}$. Notice that by choosing the value of ${\cal S}$ one defines what has to be considered as DPS (all events with $b>b_{min}$, corresponding to $|\bold P_t-\bold Q_t|<{\cal S}$) and what has to be considered as SPS (all events with $b<b_{min}$, namely $|\bold P_t-\bold Q_t|>{\cal S}$), the physical (${\cal S}$ independent) cross section being the sum of the SPS and DPS cross sections\cite{Diehl:2017kgu}.

The picture of the DPS interaction obtained in this way is the simplest extension, of the picture described in the previous sub-section, to the case where the cross section depends explicitly on $\bold P_t$ and $\bold Q_t$ and the DPDs are singular at small $b$. Interestingly, as a result of the singular behaviour of the DPDs, a limiting transverse distance separating double from single parton collisions needs to be introduced, $1/{\cal S}$, which, as shown by the expression of the term $C_1$ in Eq.(\ref{TDC}), generates a logarithmic increase of the population of interacting parton pairs at short relative transverse distances.

 \section{Concluding Discussion}

All considerations above do not include the evolution of partonic population, induced by the scale of the hard interactions ${\cal Q}^2$. Without including evolution, either in ${\cal Q}^2$ or in the relative transverse distance $b$, $\bold P_t$ and $\bold Q_t$ are both of order $1/R$ and the initial partonic populations are thus evaluated at the scale $(1/R)^2$. Evolution allows the c.m. momenta of the two interactions, $\bold P_t$ and $\bold Q_t$, assuming much larger values. When looking for DPS, one typically selects configurations where $\bold P_t$ and $\bold Q_t$ are of the order of the transverse c.m. momenta, of the final state partons with large $p_t$ observed in $2\to2$ hard parton processes, which might be a sensible choice also for the scale ${\cal S}$.  

To have some idea on possible expectations from a comprehensive study of the two-scale evolution problem, we make a few qualitative considerations along with simplifying assumptions. The main simplifying assumption being that the two evolutions, in ${\cal Q}^2$ and in $|\bold P_t-\bold Q_t|^2$, are independent one from another. Considering moreover that in DPS the hard process is disconnected in two almost independent hard subprocesses, we further assume that all evolution dynamics, which generates the initial momenta $\bold a_{i,t}$, $\bold a_{i,t}'$, can be included in the upper part of the diagram in Fig.\ref{DPS1} and all evolution dynamics, which generates the initial momenta $\bold c_{i,t}$, $\bold c_{i,t}'$, can be included in the lower part of the diagram in Fig.\ref{DPS1}. In this way all steps, leading to expression (17) in
the previous subsection, can be repeated and the conclusion that $|\bold P_t-\bold Q_t|$ and $b$ are conjugate
variables follows. Having selected the c.m. transverse momenta of the two hard interactions such that $|\bold P_t-\bold Q_t|<{\cal S}$, the corresponding range of the relative transverse distance $b$ runs from a minimum not smaller than $1/{\cal S}$, when $\bold P_t$ and $\bold Q_t$ are sizeably different in moduli and/or with a
large relative angle, to a considerably larger maximum, which may reach also values of ${\cal O}(1/R)$ when $\bold P_t$ and $\bold Q_t$ are close in moduli and with
a small relative angle.

Interacting partons may be either generated by independent evolution or by short distance
dynamics, whose dominant term is parton splitting. Independent evolution generates initial
state configurations with c.m. transverse momenta $\bold P_t$ and $\bold Q_t$ randomly distributed in their
relative angle and as a function of the relative transverse distance $b$. The initial flux of interacting parton pairs, generated by independent evolution,
would thus depend on the absolute values $|\bold P_t|$ and $|\bold Q_t|$. When the source of initial state partons
is short distance dynamics, the initial flux of interacting parton pairs grows, on the contrary, at small relative transverse distances, namely
when the angle between $\bold P_t$ and $\bold Q_t$ increases.
One my thus have a direct indication on the relative importance of splitting versus independent
evolution, by comparing the rate of DPS, in events where $\bold P_t$ and $\bold Q_t$ are parallel, with the rate
of DPS in events where $\bold P_t$ and $\bold Q_t$ lie back to back. By measuring the differences of the DPS
rate of events in the two configurations, for different choices of $|\bold P_t|$ and $|\bold Q_t|$, one would thus
obtain a direct indication of the relative importance of the two evolution mechanisms in different
kinematical configurations.

One may therefore expect to be able to learn a lot form the study of DPS as a function of the transverse centres of mass of the two hard collisions. On the other hand, while the indications from the simple case discussed here above look encouraging, an exhaustive analysis of the problem is presumably a topic for a whole research project. As already pointed out by Diehl, Gaunt and Schönwald\cite{Diehl:2017kgu}, the separation between DPS and SPS depends on the arbitrary choice of a scale (in the present note ${\cal S}$) while only the sum of the two cross sections is choice-independent. For a comprehensive study of the problem one needs therefore to include also SPS in the discussion. In addition, having to deal with two very different scales, ${\cal Q}$ and $|\bold P_t-\bold Q_t|$, where the latter scale can range from values of ${\cal O}({\cal S})$ to values of ${\cal O}(1/R)$, an exhaustive analysis of the problem requires the (non trivial) derivation of the corresponding DPD evolution equation or, at least, obtaining a viable and reasonably good approximation of it. 

As a final comment, $|\bold P_t|$ and $|\bold Q_t|$ have been already measured in several experimental studies of DPS, e.g \cite{Abazov:2011rd,Abazov:2014fha,Lincoln:2016fgq}. An analysis of the data, according with the indications here above, looks therefore already possible, at least to a certain extent. It would thus be very interesting to test if and to what degree the expectations sketched in the present note correspond to the actual experimental evidence.

\acknowledgements
\noindent
We gratefully acknowledge Markus Diehl and Jonathan Gaunt for a very useful email exchange.

\end{document}